\documentclass[10pt]{article}
\usepackage[margin=1in]{geometry}
\usepackage{authblk}
\usepackage{amsmath,amssymb,amsthm}
\usepackage{booktabs}
\usepackage{changepage}
\usepackage[colorlinks=true,linkcolor=blue,citecolor=blue]{hyperref}
\usepackage{threeparttable}
\newtheorem{proposition}{Proposition}

\theoremstyle{definition}

\title{The Missile and the Toll\\[4pt]
\large Two-Way Flow Through the Strait of Hormuz}
\author{Bernhard K Meister\thanks{bernhard.k.meister@gmail.com}}
\date{\today}
\begin{document}
\maketitle

\begin{abstract} 
\noindent 
Tolls for safe passage and attacks on shipping are usually treated as mutually exclusive: either ships pay or attacks happen. We argue they are complementary. In a Stackelberg escalation game for the Strait of Hormuz, the enforcer chooses an attack intensity and a toll schedule; the response is to pay, cross exposed, wait, reroute, or exit. Attacks create the willingness to pay; tolls harvest it. The model predicts asymmetry across trade forms: a toll where the cargo owner can pay and the vessel owner will consent, and none where either fails. Any such payment is clandestine, but the model's implications are testable against freight rates, insurance premia, and incident reports.  
\end{abstract}

\section{Introduction}\label{sec:intro}
\noindent 
Trade through a maritime chokepoint under coercive attack does not respond
uniformly. Some flows continue, reorganized and repriced; others stop, or
wait beyond the hazard. 
    Standard readings predict that trade form does not matter: if the hazard is military, impairment should track exposure; if it is political, it should track flags and alliances.
The pattern that motivates this paper, the same hazard producing different outcomes across trade forms, fits neither prediction.

\noindent 
The 2026 Strait of Hormuz episode supplies the instance. As of
October 2026, crude exports from the Gulf have resumed in
substantial volume~\cite{asharq2026flows,agbi2026shuttle}\footnote{
Regular transit commentary also appears under  \#TankerTrackers (TankersTrackers.com, Inc.) and \#live\_ais\_com (Live-AIS.com) on X.}; bulk traffic has firmed---non-Iranian-linked
bulker transits reached close to 60 in September~\cite{lloydslist2026bulkers}---while
containerized imports into the Gulf remain substantially impaired, with liner
services curtailed at Gulf ports and moved to transshipment hubs beyond
the corridor~\cite{lloydslistintel2026}. One chokepoint, one hazard, and apparently different outcomes
across the three trade forms.\footnote{All episode-specific quantities are as of October
2026 and provisional; data providers disagree on levels, and no mechanism
claim of the paper depends on them. The apparent sorting is therefore an explanandum, not
an established fact. Section~\ref{sec:evidence} mentions `classification' as an outstanding task.}

\noindent 
A flow is taxable\footnote{`Taxable' is used in the neutral sense of a flow capable of bearing a proportional charge, without implying a recognized fiscal authority. The word names the form of the charge --- a fraction of cargo value --- not the legitimacy of the enforcer.}, in the sense used here, when it is collectable. The property is organizational, not geographic; the definition is set out in Section~\ref{sec:model}.

\noindent 
The pattern is not naturally summarized by exposure alone. Containerized
imports appear more impaired than some bulk and hydrocarbon flows traversing the
same corridor. Nor does perishability sort the flows: food and medicine
\footnote{Truck and air capacity has increased during the conflict but remains a small substitute for pre-conflict maritime capacity. A single large container ship carries thousands of truckloads — more than the world's entire freighter fleet moves in one load.}
arrive on a consumption clock that cannot wait out a crisis, while crude also
faces storage, shut-in, and scheduling constraints. The useful question is
therefore not simply which cargo is more desperate, but which trade form can
reorganize around a counterparty able to transact for passage. Political
exemptions and heterogeneous risk tolerance remain competing explanations;
the empirical section treats organizational sorting as evidence about
feasibility, not by itself as evidence that a clandestine payment occurred.

\noindent 
The paper's aim is to decipher the actions of cargo owners, which requires
understanding vessel owners' interests as well. 
A voyage under a toll requires two parties, the \emph{cargo owner who pays} and the \emph{vessel owner who sails}, sanctions attach to both, from different sides. The joint structure is formalized in Section 2.
The
hundreds of ships trapped in the Persian Gulf since the conflict began are
the limiting case: ownership on both sides that precludes either paying or
hazarding the crossing, the ship deteriorating as it waits\footnote{Many
trapped vessels carry Constructive Total Loss cover, under which a ship
whose recovery is judged uneconomic is written off and the insurer pays out
the hull value. Prolonged waiting therefore converts the ship into a claim.
Legal wrangling is often part of deciding the validity of a claim.}.

\noindent 
\noindent
Applied to the Gulf, the two conditions of Section~2---a payment interface that can authorize the payment, and a vessel owner who can sail under it---produce a clear sorting: crude satisfies both, container shipping fails both, and bulk lies between, sorting within itself.


\noindent 
The paper is a mechanism-discrimination paper. We ask, \emph{suppose}
selective passage through the Strait reflects an informal protection
mechanism, what patterns must follow, and which observed patterns would
distinguish that mechanism from indiscriminate hazard or political
exemption? Three claims organize the answer. First, attacks and protection
payments can be complements rather than substitutes: some violence
establishes the hazard that makes protection valuable. Second, the
ability to extract a payment depends on collectability, not merely on
willingness to pay. Third, Gulf trade forms differ sharply in
collectability because their payment interfaces and carrier
substitutability differ; the informal fleet is an adaptation that makes
the crude export side toll-compatible. The Danish Sound serves as the
historical benchmark.
Section~\ref{sec:model} sets out the minimal model;
Section~\ref{sec:directions} discusses the two trade directions;
Section~\ref{sec:sound} describes the Sound; Section~\ref{sec:evidence} introduces some
tentative predictions and evidence; Section~\ref{sec:conclusion} concludes. The appendices contain a discussion of the indifference frontier, some illustrative  bounds, and a provisional classification protocol.

\noindent 

\section{A Minimal Model of Coercion and Collectability}\label{sec:model}

\subsection{Coercion at a chokepoint}
\noindent 
An enforcer controls a hazard at a chokepoint by choosing an attack intensity $\lambda$ and a toll\footnote{The word `toll' refers throughout to a counterfactual object, not to an observed transaction.} schedule $B$.
Each transiting party is a follower: given $(\lambda, B)$, it chooses
among four states of the voyage: pay and cross protected; cross without
paying, absorbing the hazard; wait outside the corridor; or reroute.
 All four
presuppose a live trade.  When the trade is unwound---cargo
re-sold, contract cancelled, the buyer replaced---the outcome is exit. Paying competes with three costed alternatives—crossing exposed, waiting, substituting the route or carrier—and the cheapest sets the ceiling.
Waiting keeps the option to sail and pays $D$ for holding it; exit surrenders the option and dissolves the ceiling.
Write the  combined hull-and-cargo value as
 $V_0$\footnote{$V_0$ is the \emph{laden-leg} value, hull plus cargo. The unit of decision is the round trip: a tanker discharging outside the Gulf must return, usually in ballast, at a hull-only value roughly one-third of $V_0$ under the calibration below. A toll charged on the laden crossing is therefore economically a charge on the round trip. If the toll is levied on both legs, the relevant ceiling is roughly $4/3$ of the laden-leg ceiling, since the ballast leg is also exposed.}, the mean loss fraction given a successful strike as $\bar q$, the
retained (uninsured) share as $\alpha$\footnote{$\alpha$ captures the cargo owner's retained exposure to a
strike. The vessel owner's exposure---the hull, plus the institutional
cost to the owner of any payment---generates a separate inequality of
the same form. 
}, and the attack success probability as $p_{\text{hit}}$, so the effective
hazard is $p_{\text{hit}}\lambda$.
The follower pays only if the toll, net of the institutional cost of
paying $L$ (sanctions exposure, lost cover, lost finance), does not exceed
the \emph{cheapest} outside option:
\begin{equation}
B \;\le\; \min\Big\{\;
\underbrace{\alpha\, p_{\mathrm{hit}}\, \lambda\, \bar q\, V_0}_
{\text{cross exposed}},\;\;
\underbrace{D}_{\text{wait}},\;\;
\underbrace{A}_{\text{reroute or substitute}}\;\Big\}\;-\;L,
\label{eq:ceiling}
\end{equation}
where $D$ is the cost of deferring the voyage---expenses during the
delay and the delayed arrival of the voyage margin, over the expected
wait---and $A = \min\{C_{\mathrm{alt}}, P\}$ is the cheaper of the
alternative route and exempt-tonnage substitution, all evaluated per
round trip.

\noindent 
Equation (1) is written from the cargo owner's perspective: the payment
decision and the toll cost are the cargo owner's. The vessel owner
faces a separate inequality---cleared not by the toll but by the
freight, insurance, and charter terms---and the voyage requires both
to be satisfied. Section 3 works out qualitatively which pairings
clear both.


\noindent 
Three remarks on \eqref{eq:ceiling}. First, the ceiling is the
\emph{minimum} of the outside-option costs, not their sum and not their
maximum: a follower abandons paying at its cheapest alternative, so the
toll is bounded by the cheapest escape. Second, the ceiling rises with the threat: the value of a toll is the danger it removes. 
The ceiling is one
point on a larger object: the follower's indifference frontier between
paying and bearing the hazard, further developed in Appendix~A. 
Attacks and
tolls are therefore complements---some violence establishes the hazard that
makes protection valuable, and a threat never realized decays in the
market's estimate. 
This is the protection-payment logic in its classical
form \cite{schelling1960strategy,lane1979profits, reuter1987racket,skaperdas1992cooperation,gambetta1993sicilian}:
the enforcer sells relief from a hazard it also produces.
Third, the ceiling in \eqref{eq:ceiling} is conditional on the voyage
occurring, and the voyage requires both parties---either can exit
independently. The vessel owner exits when the round-trip margin falls
below the vessel's next-best use; the cargo owner when the trade is
surrendered. The exits interact: vessels exiting activate the
substitution branch of \eqref{eq:ceiling}; cargoes exiting shrink the
enforcer's base beyond the reach of any schedule. Waiting degrades into
exit on either side as the storage clock runs. For low-margin or short-clock
cargo, these constraints bind before the ceiling does.


\begin{proposition}[Ceiling and targeting]\label{prop:ceiling}
The ceiling in \eqref{eq:ceiling} is non-decreasing in the enforcer's
hazard. When the cross-exposed branch binds, $B^{\max} = \alpha
p_{\mathrm{hit}}\lambda\bar q V_0 - L$, so the maximum toll is increasing in the enforcer's own targeting  intensity $\lambda$:
price discrimination across counterparties is manufactured by targeting, not
inherited from geography.
\end{proposition}

\begin{proof}
The cross-exposed cost $\alpha p_{\mathrm{hit}}\lambda\bar q V_0$ is
strictly increasing in $\lambda$, and the remaining branches are
non-decreasing in it---congestion raises the wait cost, and neither
the alternative route nor exempt substitution depreciates with the
hazard---so the minimum is non-decreasing; where the cross-exposed
branch binds, the ceiling inherits the strict monotonicity. Indexing
 $\lambda$ by counterparty gives the second statement.
\end{proof}

\noindent 
Escalation is nonetheless bounded, because the enforcer's revenue is
downstream of the flow it taxes. 
Higher hazard raises the ceiling, and with it the toll per payer, but it also drives transits out of the corridor and, among those remaining, pushes some toward the non-paying branches. 
An extraction-oriented enforcer therefore has an incentive to calibrate
violence---both its intensity and its targeting---so that it enhances, or
at least preserves, the rent base\footnote{For tanker traffic this
incentive is reinforced by environmental self-harm: a terrestrial analogue
of the Kessler dynamic, in which use of a shared resource degrades the
resource itself. A large spill or burning slick can impair the corridor,
foul cooling-water or desalination intakes, and raise costs for traffic on
which the enforcer itself depends. The exact damage threshold is highly
uncertain; the relevant point is only that pollution is an additional
downside to attacks on laden tankers, not a separately modeled channel.} rather than exhausting it.
This property---revenue lies downstream of the flow---is what gives
the attack-composition test of Section~\ref{sec:evidence} its sign. 
The repeated-game
analysis is not needed for any
prediction made here\footnote{Two further boundaries of escalation, stated here only:
attacks fail with probability $1-p_{\mathrm{hit}}$, which scales the
ceiling down by $p_{\mathrm{hit}}$ and raises the effective marginal cost
of escalation; and the exempt share matters, since exempt traffic pays
nothing and the direct escalation channel scales with the non-exempt
share, vanishing as the exempt share approaches one.}.

\subsection{Collectability}
\noindent
Willingness to pay is not enough. Define a transit as collectable when two
conditions hold: there exists an identifiable payment interface---an actor
with both the authority and the ability to pay---and there exists a carrier
able to move the cargo. For both, the institutional penalties — sanctions exposure, lost cover, lost finance — are real, but must not deter the parties involved.
Collectability names this property. 

\noindent
Toll revenue is proportional to the collectable flow:
\begin{equation}\label{eq:revenue}
\text{revenue} \;\propto\; B \, Q_c,
\end{equation}
where $B$ is the toll and $Q_c$ the number of collectable
transits, i.e. those for which the two conditions hold.
Cargo value also matters, but separately, since it
constrains the toll $B$ that can be charged, not whether the flow is
collectable itself. 

\noindent
Fragmented ownership can undermine collectability, but it is not
sufficient by itself to do so. Ordinary transport systems routinely
aggregate charges across many cargo owners. What matters is whether
aggregation and payment capability coincide in the same actor, and whether
that actor can act despite the penalties. When those
capacities do not meet, coercion produces interruption or rerouting rather
than a clandestine price.

\noindent
The same physical hazard can
produce extraction in one trade organization and interruption in another,
because the  feasible action set varies. 

\noindent
What does coercion look like when a flow is not collectable? Not a price,
and not nothing. On the collectable leg, the hazard produces revenue; on
the uncollectable leg, it produces deadweight

\section{Two Directions of the Same Corridor}\label{sec:directions}
\noindent
Different trade flows pass through the Strait of Hormuz, and whether a toll
can be paid depends on the type of cargo owner, the type of vessel owner,
and the type of cargo. The \emph{cargo owner} authorizes the payment;
the \emph{vessel owner} consents to the voyage; and the type of cargo
determines which combination is sensible. Three cargo-owner types
are considered: a national oil company (NOC), a non-sanctioned private trader, and a
sanctioned trader. Three vessel-owner types are considered: sovereign
tonnage, standard commercial tonnage, and informal tonnage\footnote{Terminology: this paper uses `informal tonnage' as the
umbrella for vessels operating outside the `formal' insurance,
financing, and classification system---typically older hulls under
hidden ownership. The subsets differ in visibility, not in kind:
`grey fleet'---still trackable, compliance thin but extant, some
institutional exposure retained; `dark fleet'---largely untrackable,
institutional exposure exhausted; `shadow fleet' covering both. The
categories are broad classifications, not exact definitions; the
paper distinguishes subsets only where the mechanism requires it.}
outside the formal system. Four cargo types are considered: crude oil,
LNG, bulk, and containerized cargo. The sections below take each in turn
and ask which pairing of cargo owner and vessel owner can satisfy the
joint condition.

\subsection{Crude Exports}

\noindent
A very large crude carrier (VLCC) cargo is a single consignment of roughly
two million barrels\footnote{The cargo of thirty
VLCCs would bring average concentration to roughly one part per million—the order of magnitude at which reverse-osmosis desalination is affected. The figure is a scale calculation only, since oil does not dissolve uniformly, and concentrations near a spill dominate any basin average.}. For crude, the cargo and vessel ownership may coincide---a NOC can own both---but they are usually distinct. 

\noindent
Cargo owners come in three types. A NOC is
sovereign: the institutions that would penalize a private counterparty
have no simple standing against it, and its exposure is mainly political. A non-sanctioned trading house is inside the formal system and
bears the full institutional penalty if the payment becomes attributable.
A sanctioned trader is  outside the system, and often acts
through shell companies whose connection to the designated entity is
difficult to establish; enforcement against such arrangements is  
uncertain.

\noindent
Vessel owners also come in three types. Standard commercial tonnage is
club-insured, listed, and dollar-settling; its owner bears the full
institutional penalty. NOC-owned tonnage is controlled by a sovereign, so
its exposure is the same as a sovereign cargo owner's. Informal tonnage is already outside the formal system---no club cover, no
dollar clearing, often already sanctioned---and a payment adds nothing to
its exposure because there is nothing left to add\footnote{The claim concerns the marginal exposure of the payment, not of the voyage. Coalition navies have seized tankers tied to the current conflict, not sanctioned tonnage as a class; the hazard is targeted at particular operations, not categorical. The enforcer does not control it, and the informal owner's consent is priced against it.}.

\noindent
The enforcer in the current episode is designated, so a payment to it is,
for any party inside the formal system, a sanctions violation. A toll is
feasible only when neither the cargo owner nor the vessel owner is
deterred by the exposure the payment would create.  A sovereign cargo owner on standard tonnage can pay when the payment does not attach to the vessel. When it would, and the cargo owner cannot shield the vessel owner sufficiently, then the charter migrates to NOC-owned or informal tonnage. 


\noindent

\noindent
The deterrence each party faces has the same structure. 
The three determinants of deterrence are
the probability of attribution, the penalty if
attribution succeeds, and the party's attitude toward the resulting risk.
The probability of attribution depends on how opaque the payment can be and
on the enforcement effort directed at it. The penalty depends on the
party's status: for example, a sanctioned owner has limited additional risk,  a
non-sanctioned private owner is exposed to sanctions, its officers to
personal liability, and a sovereign to political fall-out. Attitude
toward risk is what makes a possible cost acceptable to one party and not
another, and the paper treats it as idiosyncratic. 

\noindent The structure above suggests payment feasibility.
One category sits outside the above structure: politically exempt traffic, meaning flags the enforcer's state has reasons to spare\cite{reuters2026china}. It is collectable in principle but priced at
zero by political arrangement: outside the paying pool not because it
cannot transact, but because it need not. Exemption is a political
analogue of the toll. 


\noindent


\noindent
As an aside, the unit of decision is the round trip, not the crossing. A tanker that
discharges outside the Gulf must return, usually in ballast, and the
cross-exposed ceiling on a ballast leg is roughly the hull value alone
--- roughly a third of the laden-leg ceiling under the calibration
below. The enforcer therefore prices the round trip at the laden-leg
ceiling, and the relevant bound is the round-trip margin rather than the
one-way cargo value. This tightens, rather than loosens, the incentive to
preserve the traffic: a toll that kills the round trip kills the rent
base.


\noindent


\noindent 


\noindent 

\noindent 

\noindent 

\noindent 

\noindent 
\subsection{LNG Exports}

\noindent
    Liquefied natural gas (LNG) is a contrasting case. LNG carriers are highly specialized, and the fleets operate  within the formal system; no meaningful informal LNG tonnage exists, as it does for crude. LNG has continued to cross the Strait, on both NOC and chartered standard vessels, but more sparingly than crude.

\subsection{Bulk: the intermediate case}
\noindent 
Collectability is a spectrum, and bulk
occupies it unevenly because bulk is not one class. Bulk exports---sulphur,
urea, limestone---share the concentrated, often state-linked ownership of
crude: one principal per consignment, direct chartering, capacity to
migrate towards tonnage outside the formal system. Their payment interface
is available. But two constraints push them short of crude. Lower value
density raises the fixed cost of a clandestine transaction relative to
cargo value, so the toll they can support is smaller. 
And these flows are less central to the state's rent base than crude is, so the state has less reason to keep them moving through a costly toll, which lowers the effective ceiling.
 Bulk exports therefore track crude with a
caveat: the interface exists, but the flow supports partial rather than
full adaptation. 
Bulk imports---grain, fertilizer, construction
materials---are usually controlled by commercial principals, 
supplemented by government agencies during the conflict. 
Commercial principals face lower value density and greater formal-finance exposure, and sit further toward the middle. 
But where the state itself becomes the importer, the follower's characteristics shift toward the crude case, and the flow moves closer to selective passage than to partial adaptation. The model therefore predicts not only that bulk is intermediate between crude and containers, but that it sorts within the category: by ownership, by actor type, and by the fiscal stakes attached to the flow.

\subsection{Container Imports: unbillable and uninsurable}

\noindent
A container vessel entering the Gulf carries the goods of thousands of consignees: food, medicine, spare parts. The liner, as the operator, is the aggregator: it can and routinely does pass port, canal, fuel, and risk charges through to customers.
The difficulty is
institutional rather than arithmetical. Global liner companies are among
the most formal actors in the system---listed, club-insured,
bank-financed, and sanctions-sensitive---so the only actor capable of
aggregating the payment is also the actor least able to make an illicit
one. 
 Individual consignees cannot  bypass that constraint simply by chartering a substitute liner network, because the same problem is faced by all the major liner networks.

\noindent
The container leg also fails on the carrier side.  Container lines aggregate cargo and vessel liability  onto themselves. The hull is its own, and a casualty makes the liner the defendant
to every consignee at once: a liability spread across thousands of bills
of lading\footnote{The consignees separately insure the contents against loss,
but that cover does not relieve the liner: once a cargo insurer has paid
its consignee, it subrogates to the consignee's rights and claims
against the carrier under the bill of lading, within the carrier's
Hague--Visby and global-limitation (LLMC) caps. The liner's own exposure
therefore remains---its capped liability to every consignee on board,
plus the litigation risk that a war-casualty defense fails---and
war-risk P\&I, not ordinary P\&I, is the cover that responds. The
consignee's cargo cover and the liner's liability cover are separate
purchases with separate buyers; the freight rate contains the second,
not the first.}.  The value at risk on a fully laden ship is on
the order of a billion dollars, though the liner's own legal liability
for the goods is capped and far lower. Both war covers the voyage
requires---hull war risk and war-risk P\&I---are reported to be
withdrawn for this class rather than repriced~\cite{axco2026}, and the
outcome is route cancellation.
Crossing exposed is then not viable for a liner, and the enforcer cannot sell relief from a hazard the class will not face.

\noindent
    The payment interface therefore fails, and the carrier side fails with it. Both conditions of collectability are now closed, and waiting, rerouting or exit become the only options.
The corridor empties
of the class that cannot be invoiced, so its scarcity among reported
attack targets is a matter of absence, not of restraint (a claim to be
checked against the transit volume, not the incident record).




\section{The Sound as the Historical Benchmark}\label{sec:sound}
\noindent 
The Danish Sound Dues\footnote{More detailed data is available in the digitized
Sound Toll Registers~\cite{str}.} (1429--1857) provide a historical benchmark: a chokepoint toll
that operated on \emph{both} directions of trade, at a posted ad valorem
schedule on the order of 1--2\% of cargo value, anchored by the cost of
the next-best route (the Belts), for four centuries, with little continuing violence once enforcement was regularized\footnote{It was contested early on, by the Hanseatic cities and in
Swedish and Dutch wars. The British Navy, preeminent for much of the period, chose not to abolish the Dues by force. Baltic imports of masts and hemp were essential to the fleet, and a war that closed the Sound would have been strategically catastrophic. The Dues were a commercial cost Britain paid for centuries, and were ended in 1857 by redemption rather than conquest.}. The Sound, the Gulf export leg, and the Gulf import leg are the same mechanism under different institutional constraints: the security of the collection right, the legibility of enforcement, and who can transact for the cargo. Table~\ref{tab:sound} sets out the comparison.

\noindent 
First, collection mechanics. The Sound had a fortress at a
four-kilometre narrows and a recognized customs interface through which a
vessel, master, or merchant agent could be assessed against declared cargo.
The important fact is not that every item aboard had one ultimate owner, but
that the regime had a lawful payment interface capable of aggregating the
charge and passing it onward. Whatever passed could therefore be billed in
both directions. Gulf crude resembles the Sound not because ownership is
literally identical, but because a small number of principals control carriage
and can reorganize it.

\noindent 
Second, violence and legibility. Danish enforcement was architectural
and, over time, buttressed by treaty; once the collection right was recognized,
the threat required fewer demonstrations. The Sound is therefore a useful
limit case for attack/toll complementarity: visible and institutionalized
enforcement can support collection with little continuing violence. A
clandestine and contested enforcer lacks that advantage and may need repeated
demonstrations of capability. This does not make attack rates a unique index of
contestation---military and political motives can generate the same events---but
it explains why continuing violence is not inconsistent with an extraction
mechanism.

\noindent 
Third, trade organization. The Sound shows that fragmented commerce is
not inherently untaxable when a recognized intermediary and collection
procedure aggregate the charge. The modern liner system also has an aggregator,
but its institutional constraints are different: the liner can pass through a
legal surcharge but cannot readily substitute into a clandestine payment
relationship. The contrast therefore sharpens rather than weakens the
collectability argument. The Sound Dues also coexisted for centuries with large
trade volumes; that fact does not identify the counterfactual welfare cost of
the toll, but it does show that a durable chokepoint charge can remain below the
outside option and preserve the flow on which it depends.

\begin{table}[h]
\begin{adjustwidth}{-2cm}{-2cm}
\footnotesize
\setlength{\tabcolsep}{3pt}
\begin{threeparttable}
\caption{One model, four settings: The physical hazard is common to all
Gulf columns; what differs is the trade form. Expected outcomes under the
protection mechanism; with the Sound acting as a benchmark.}
\label{tab:sound}
\begin{tabular*}{\linewidth}{@{\extracolsep{\fill}}
    p{3.3cm}p{3.6cm}p{3.6cm}p{4.4cm}p{3.6cm}@{}}
\toprule
 & Danish Sound & Gulf crude exports & Gulf bulk & Gulf container imports \\
\midrule
Payment interface & recognized customs interface; pays licitly & NOC or trader; payment via opaque channel & \textit{exports:} state-linked producer; payment via opaque channel\newline\textit{imports:} commercial principal or state-linked; payment via opaque channel & liner; generally cannot pay \\
Carrier substitutability & high within merchant shipping & high, in informal fleet tonnage & \textit{exports:} high, via chartered tonnage\newline\textit{imports:} moderate for commercial principals, high for state agencies & low at network scale \\
Collectable?\tnote{a} & yes & yes & \textit{exports:} yes\newline\textit{imports:} partly & no \\
Expected outcome & toll, both directions & selective passage & \textit{exports:} partial adaptation\newline\textit{imports:} partial adaptation for commercial, selective passage for state-aligned & waiting, interruption \\
\bottomrule
\end{tabular*}
\begin{tablenotes}[flushleft]
\small
\item[a]  A trade form is collectable when two conditions hold: an identifiable payment interface---an actor with the authority and the ability to pay---and a carrier able to move the cargo, with sanctions exposure, lost cover, and lost finance not deterring the parties involved.
Bulk exports satisfy both, so a toll is feasible---but the toll they can
support is smaller, because lower value density lowers the ceiling. For
the marginal parcels the ceiling falls below the fixed cost of arranging
the transaction, so those flows wait. Collectability and the ceiling are
distinct: the first determines whether a toll can exist, the second how
large it can be.
\end{tablenotes}
\end{threeparttable}
\end{adjustwidth}
\end{table}

\section{Predictions and Evidence}\label{sec:evidence}

\noindent 
Three mechanisms could generate parts of the observed pattern. Hypothesis One (H1): 
protection mechanism with selective payment and protection. Hypothesis Two (H2):  political or
military coercion without pricing. Hypothesis Three (H3): organizational adaptation to hazard, i.e. 
formal and informal carriers differ in risk tolerance, insurance, finance, and
ability to reroute even when no clandestine payment exists. 
Table~\ref{tab:control} sets out what each implies.
H3 is important
because it can generate much of the same macro sorting as H1. Price data and
fleet composition therefore establish feasibility and adaptation, not payment.
Any toll is treated as latent. We
distinguish throughout what is documented (schedules, port calls,
AIS-observable traffic composition) from what is uncertain wartime reporting
(attack attribution, rumored payments). The useful discriminators are those
that compare otherwise similar traffic and ask whether treatment differs in a
way ordinary risk tolerance cannot explain.

\noindent
The hypotheses are not mutually exclusive in generation: attacks can
reflect political and extraction motives simultaneously, and the
empirical question is whether a residual pattern consistent with
extraction---conditional sparing, treatment discontinuities at changes in
transaction capacity---remains after conditioning on conflict events and
flag-based targeting.

\begin{table}[h]
\centering
\small
\setlength{\tabcolsep}{4pt}
\begin{threeparttable}
\caption{ Identifying implications, by hypothesis: Each row is an empirical pattern; each column is a hypothesis; entries mark whether the hypothesis predicts the pattern, permits it, or does not specifically speak to it.}
\label{tab:control}
\begin{tabular}{@{}p{9cm}p{2.1cm}p{2.1cm}p{3.8cm}@{}}
\toprule
Implication & H1 & H2 & H3 \\
\midrule
Sorting towards opaque tankers and away from formal liners & predicted & possible & predicted \\
Import waiting and rerouting by trade organization & predicted & possible & predicted \\
Conditional sparing of otherwise similar traffic associated with protected or payment-capable status & predicted & possible & not specifically predicted \\
Transaction evidence & predicted\tnote{a} & no & no \\
\bottomrule
\end{tabular}
\begin{tablenotes}[flushleft]
\small
\item[a] `Predicted' means the hypothesis implies the pattern, not that it is confirmed by available sources.
\end{tablenotes}
\end{threeparttable}
\end{table}
\noindent

\paragraph{1. The sorting across and within trade forms.} The premise of Section~\ref{sec:intro} is the first object of classification: whether outcomes differ systematically by trade form.
The reported bulker series~\cite{lloydslist2026bulkers} (around sixty non-Iranian-linked transits in September, mixing laden and ballast legs so that the round-trip count cannot be recovered) is useful precisely because it anchors the bulk side of a contrast between a  bulk trade that continues and container traffic  that has all but disappeared from the corridor.       The series is aggregate: it filters by linkage status, not by commodity, direction, actor type, or fiscal stakes, so it cannot test the within-bulk sorting the model predicts.
 That test requires parcel-level data---commodity, ownership, charter arrangement, direction---from cargo trackers or customs records, set out as a protocol in Appendix~\ref{app:measure}.

\noindent    
If the bulk–container contrast fails eventually in the data, the puzzle dissolves. If it holds, it establishes
organizational adaptation but does not yet distinguish H1 from H3.

\noindent    
  The attack record cannot perform this discrimination: if the class that cannot be invoiced does not present itself, the observed attacks will look like exposure-tracking even when the mechanism is organizational. 
  The discriminating evidence is therefore in the transit volumes: whether container traffic recovers when the hazard falls, and whether it recovers proportionally or remains suppressed by the payment constraint.

\paragraph{2. Composition of surviving traffic.} Under both H1 and H3,
informal tanker tonnage can be overrepresented among
surviving transits while formal liners withdraw. AIS\footnote{It is not uncommon for ships traveling in the region to switch of their AIS for part of the journey. AIS spoofing is also common.} composition by flag, age,
ownership, and insurance class is therefore evidence for institutional
segmentation, not direct evidence for payment. Its value is as a conditioning
set for the sharper tests below.

\paragraph{3. Conditional treatment within comparable traffic.} The
protection hypothesis earns empirical content only where it predicts different
treatment among otherwise similar vessels. Because revenue is downstream of the
flow it taxes, an extraction-oriented enforcer should preserve the traffic it
monetizes. The relevant test is therefore conditional: among vessels with
similar route, cargo, flag exposure, escort, and observable risk, does attack or
interference vary discontinuously with independently observed eligibility,
counterparty status, or changes in the institutional ability to transact—changes visible in registries as sales, reflags, and P\&I withdrawals? Indiscriminate attacks on the supposedly monetized flow count against H1;
selective survival explained entirely by age, insurance, or risk tolerance does
not identify it. Finer claims about targeting intent, schedule revisions, or
placement on the hull require assumptions the incident data cannot bear and are
not made.

\paragraph{4. Waiting and direct evidence.} For the export leg, waiting is anchorage dwell inside the risk perimeter, tracked by AIS where available. For the import leg, waiting appears as port-call omission and surcharge schedules at hubs outside the corridor, both observable in published liner schedules. These are clean measures of adaptation and deadweight
loss, but not by themselves of payment. Strong identification of H1 would come
from treatment discontinuities tied to independently observed eligibility or,
best of all, direct transaction evidence. 

\paragraph{The trapped fleet.} The trapped fleet is the both-deterred cell
made visible: vessels whose cargo and ownership jointly preclude payment
and hazard. Under constructive-total-loss cover, prolonged waiting converts
a ship into an insurance claim, so the eventual hull-claim record will be an observable
of the cell. Its growth maps the boundary of the paying pool from below.

\paragraph{Hazard triangulation.} Attack counts, war-risk premia, freight
differentials, and traffic composition are all manifestations of one
latent hazard and can each support an independent estimate of it:
incident-based (from UKMTO counts and voyage totals), market-based (from
premia net of base rates), and behavioral (from rerouting and waiting).
No single data series is sufficient, and the value of the exercise is
negative: sustained premia through quiet periods are inconsistent with
salience-maintenance under H1 and point instead to scarcity rent; attack
clusters that leave premia unmoved are inconsistent with any
hazard-pricing reading, H1 or H3. 
The protection hypothesis interprets a latent variable and cannot be tested directly.  The data instead constrains what is plausible.

\section{Conclusion}\label{sec:conclusion}
\noindent 
Three claims. Attacks and protection payments can be complements rather
than substitutes: some violence establishes the hazard that makes protection
valuable. The ability to extract a payment depends on collectability, not
merely on willingness to pay. Gulf trade forms differ in
collectability because their payment interfaces, institutional constraints, and
ability to substitute carriage differ; informal tonnage is an
adaptation that makes the crude export side more sanction- and toll-compatible. 

\noindent 
The general result is that a coercive protection market has an
organizational boundary determined by collectability, which is a property of
trade organization rather than vulnerability alone. 
The boundary itself is set pairwise. Cargo owners run from the
sanctioned shell company with nothing left to lose to the ESG-anchored
European major with everything to lose; vessel owners from dark
tonnage to the institutionalized liner. A toll clears a pairing only
if it clears both parties, so each flow's taxability is set by the
more deterred of its two members: the minimum of the pairing, not the
average of the trade. The tolerant extremes meet and can transact; the
intolerant ones do not sail at all; and the corridor's sorting is the
sorting of these pairwise minima.
The Strait of Hormuz
supplies the application and the Sound a benchmark. The current directional and
institutional asymmetries are evidence for organizational segmentation; whether
they also imply a protection market depends on the within-class and direct tests
set out in Section~\ref{sec:evidence}.

\noindent

\noindent
If an informal protection regime operates in the Gulf, it resembles the Sound's contested imposition more than its mature, treaty-regularized form.
The model does not determine the resulting long-run order. The contribution here is narrower: once coercion makes passage scarce, trade organization helps determine which flows can be priced, which can adapt, and which simply stop. 

\noindent 
Attacks on shipping are frequent and the situation is fluid.
Will the corridor settle toward the Sound's mature phase—a posted charge and low violence, whatever its legality—or revert to the status quo ante?

\appendix

\section{Indifference Frontier \& Credibility}\label{app:model}

\noindent
\noindent

\noindent 
\noindent 
Two self-contained complements to the model of Section 2 are
provided: the follower's indifference frontier, drawn from the geometry of
(1), and the repeated-game condition under which the enforcer spares
payers rather than collects and seizes.

\noindent
\paragraph{The indifference frontier.} The ceiling of (1) is the value of
the follower's indifference frontier at the prevailing hazard. Each
branch of the minimum in (1) generates one boundary of that frontier. In $(B,\lambda)$ space, the boundary generated by the pay-versus-cross-exposed branch is
\[
B + L = \alpha p_{hit}\lambda\bar q V_0, \tag{3}
\]
a line with slope $\alpha p_{hit}\bar q V_0$ (dollars of toll per
unit of hazard, per voyage) and intercept $-L$, with $\alpha$ and $L$ the follower's retained share and institutional cost from Section 2. 
Section 3 gave a family of such lines: steep for the
uninsured owner ($\alpha\approx 1$), flat for the club-insured; near
the origin for the sanctioned trader ($L$ small), far left for the
listed one. Each remaining branch of (1) generates a boundary of the
same kind 
and the follower's true frontier is the lower envelope
of the family\footnote{    The vessel owner's consent conditions add constraints of the same kind, so the frontier of a transacting pair is the lower envelope of both parties' boundaries.
 }. 
The enforcer's choice of $(B,\lambda)$ is a point in this plane: price
discrimination by targeting (Proposition 1) is relocating the point onto
a chosen line in the family. Where the envelope never rises above
zero---the container case---no one pays at any hazard, and coercion
produces interruption rather than a price. The frontier is the demand side of the
protection market; the supply side---how far the rational enforcer
escalates---is bounded  by the traffic that escalation
itself destroys.

\paragraph{Credibility.} 
\noindent
A toll buys passage on one crossing; nothing binds the
enforcer after payment, and the one-shot temptation to collect and
seize is present every time. Let $\delta$ be the enforcer's discount factor per period, where a period is one round trip for which a toll is due; $\bar B$ and
$\gamma V_0$ are on the same period basis. 
Cooperation (sparing payers) is
sustainable iff $\delta \ge \delta^* = \gamma V_0/(\gamma V_0+\bar B)$,
the case of certain detection in the condition below. The value of a
cooperative stream beginning next period is $\delta\bar B/(1-\delta)$;
a deviation strategy---collect and seize, repeating until
detection ends the stream, an event that occurs each period with
probability $p$---is worth
$\delta(\bar B+\gamma V_0)/[1-\delta(1-p)]$. The
common factor $\delta$ cancels. $\bar B$ is the average toll and
 $\gamma V_0$ the enforcer's share of seizure value, not the victim's
loss. With detection probability $p$, the condition becomes $\delta
\ge \gamma V_0/(\gamma V_0 + p\bar B)$: misattribution tightens the
requirement, and a large prize tightens it further: the admissible
seizure-to-toll ratio is $p\,\delta/(1-\delta)$. What constrains cooperation are attribution and the prize, not patience. For an
organized enforcer with a horizon beyond a few weeks, sparing payers
is self-interested when the toll is large relative to the prize; at a
\$5M toll against a \$50--150M seizure share the required horizon
stretches to months.

 \section{Illustrative Bounds}\label{app:calib}

\noindent
The main text treats the toll as latent. For illustration only---no
prediction depends on it---observable data bound a toll as follows.
Write $V_0=\$300$M, with hull $\sim\$100$M and cargo $\sim\$200$M. When
the cross-exposed branch of \eqref{eq:ceiling} binds, the toll ceiling is the
expected-loss component alone: the part of the market price of
unprotected transit that compensates the hazard rather than the risk
premium, frictions, or scarcity rent.

\noindent 
The war-risk premium moved from a pre-event 0.05--0.2\% of hull value,
to 1.5--3.0\% during early escalation, to 7.5--10\% at the episode's
earlier peak \cite{axco2026,spglobal2026marsh}; quotes following the
early-October spike will likely set new highs\footnote{Reported broker ranges, not observed
transaction prices; actual premia are negotiated privately per voyage
and vessel.}. Applied to a \$100M hull, the widest reported band bounds the
hull-side expected loss from above at \$7.5--10M per voyage\footnote{Throughout, $\bar q = 1$. A successful strike is assumed for simplicity to destroy the insured
value completely.}.

\noindent
If the enforcer captures roughly half the ceiling\footnote{Half is a convention—the model delivers the ceiling, not the fraction. At the ceiling the payer sits on the indifference frontier, exactly indifferent between paying and bearing the hazard, and any friction, transaction cost, or doubt that the payment binds the enforcer tips that indifference toward refusal. An enforcer that harvests willingness rather than exhausts it therefore leaves a margin; the bounds scale linearly in the fraction, so any other convention can be substituted.}, then a reasonable toll on hull alone is  \$3.75--5M per voyage, or 1.2--1.7\%
of $V_0$ and 1.9--2.5\% of cargo value---a lower bound on the
combined toll, since the hull is one component of $V_0$. If the cargo
faces the same risk per dollar as the hull, the ceiling triples with
 $V_0$, and the illustrative operational toll is 3.75--5\% of $V_0$ and 5.6--7.5\% of cargo
value. 

\noindent
\noindent
The attack record enters as a direction, not yet a calculable level. 
The raw attack count of about one vessel per day in the most recent
reporting week is insufficient for a simple hazard estimate, since
attack rates vary by vessel class; producing one requires attacks on
a class over transits of that class.  The October spike 
implies a higher hazard rate. By the monotonicity of
Proposition~1 the  toll  will jump above the earlier calibrated ranges, in an amount the future premia will reveal.


\paragraph{Minimum cargo value.} 
A clandestine transaction carries on both sides of the exchange fixed costs: arranging, monitoring, enforcing payment. These are participation costs, not toll bounds. When the transaction cannot cover them, no toll is paid. The flow takes whichever of the other branches is best: cross exposed, wait, reroute, or exit. The model does not include this fixed cost, and it cannot be estimated from the public record.
\section{Classification \& Measurement Protocols}\label{app:measure}
\noindent 
The tests of Section~\ref{sec:evidence} require data.
Port-call and port-omission series, with surcharge schedules, serve the sorting tests.
AIS-derived traffic composition, supplemented by registry and P\&I data, conditions the sharper tests. 
Anchorage dwell-time data serves the waiting measure. Attack-incident tables, from UKMTO\cite{ukmto2026} and JMIC, serve the conditional-treatment and hazard-triangulation tests.

\paragraph{Bulker transit series and the within-bulk test.} Some bulker
data is available (see, for example, Lloyd's List~\cite{lloydslist2026bulkers}),
but granularity is a problem. Testing whether bulk sorts by ownership and
value requires, for each crossing, the direction, commodity, ownership
type, and charter arrangement.
That in turn requires parcel-level data, from cargo trackers or customs
records.


\end{document}